\documentclass[twoside,fleqn,a4paper]{article}
\usepackage{espcrc2}
\usepackage{graphicx,color}
\title{
Mass determination from Constraint Effective Potential
}
\author{A. Agodi \\ 
        G. Andronico\\
        Dipartimento di Fisica dell'Universit\'a di Catania\\
        INFN Sezione di Catania}
\begin{document}
\begin{abstract}
  The Constraint Effective Potential (CEP) allows a determination of
  the mass and other quantities directly, without relying upon
  asymptotic correlator decays.  We report and discuss the results of
  some mass calculations in $\left(\lambda \Phi^4 \right)_4$, obtained
  from CEP and our improved version of CEP (ICEP).
\end{abstract}
\maketitle
\section{Introduction}
It has been shown that an Improved version (ICEP) of the Constraint
Effective Potential (CEP)\cite{fund2} reduces finite size effects in
$\left(\phi^{4}\right)_{4}$ lattice calculations\cite{myself}.  The
Constraint Effective Potential $U\left( \Omega ,\overline{\phi
    }\right)$ (where $\Omega$ is the lattice 4-volume and
$\overline{\phi }$ the VEV of the field) was defined as
\begin{eqnarray}
\label{eq:Udef}
\exp && \left( -\Omega U\left( \Omega ,\overline{\phi }\right)
\right) =\\
&& \int D\phi \, \delta \left( M\left[ \phi \right]
-\overline{\phi }\right) \exp \left( -S\left[ \phi \right] \right) \nonumber
\end{eqnarray}

being $M\left[ \phi \right] = \frac{1}{\Omega} \int \mathrm{d}^d x
\phi \left( x \right)$. With the function $W\left( \Omega ,j \right)$
of the external source $j$, defined by
\begin{eqnarray}
\label{eq:Wrel}
 \exp &&\left(\Omega W\left( \Omega ,j \right)
\right) = \\
&&\int d\overline{\phi }\exp \left[ \Omega \left(
j\overline{\phi }-U\left( \Omega ,\overline{\phi }\right) \right)
\right], \nonumber
\end{eqnarray}
the effective potential $\Gamma$ is the Legendre transform
\begin{displaymath}
 \Gamma \left( \Omega ,\overline{\phi }\right) =
\sup_j
\left( j\overline{\phi }-W\left( \Omega ,j\right) \right).
\end{displaymath}
It has been shown that
\begin{displaymath}
\lim_{\Omega \rightarrow \infty}
U\left( \Omega ,\overline{\phi }\right) =
\lim_{\Omega \rightarrow \infty}
\Gamma \left( \Omega ,\overline{\phi }\right).
\end{displaymath}
For big enough \( \Omega  \)
\begin{equation}
\label{eq:G_CEP}
\Gamma \left( \Omega ,\overline{\phi } \right) \approx U\left(
\Omega ,\overline{\phi } \right) 
\end{equation}

We have shown\cite{myself} that  better results for the values of
\begin{displaymath}
  J=\frac{\partial \Gamma \left( \Omega ,\overline{\phi } \right)
}{\partial \overline{\phi } }
\end{displaymath}
are obtained by evaluating
(\ref{eq:Wrel}) with the saddle point method. In this way we get

\begin{equation}
\label{eq:G_ICEP}
 \Gamma \left( \Omega ,\overline{\phi } \right) =U\left( \Omega
,\overline{\phi } \right) +\frac{1}{2\Omega }\ln U''\left( \Omega
,\overline{\phi } \right) +K\left( \Omega \right) 
\end{equation}
where \( K\left( \Omega \right)  \) is $\overline{\phi
}$-independent and 
\begin{displaymath}
\lim_{\Omega \rightarrow \infty}
K\left( \Omega \right) =0.
\end{displaymath}
This is what we call Improved CEP (ICEP).\par
In the present work we present some preliminary results, as
obtained from the behavior of
\begin{displaymath}
\Gamma'=\frac{\partial \Gamma \left( \Omega ,\overline{\phi }
\right) }{\partial \overline{\phi } } 
\end{displaymath}
and
\begin{displaymath}
\Gamma''=\frac{\partial ^{2}\Gamma \left( \Omega ,\overline{\phi }
\right) }{\partial \overline{\phi } ^{2}} 
\end{displaymath}
on a 16$^4$ lattice.

\section{CEP}
From the assumption (\ref{eq:G_CEP}) whose reliability was checked in \cite{myself} it
follows
\begin{eqnarray*}
\Gamma''&=& U''\left( \Omega ,\varphi \right) = \\
&& \left\langle
V''\right\rangle _{\overline{\phi }}-\Omega \left\langle \left(
V'-\left\langle V'\right\rangle _{\overline{\phi }}\right)
^{2}\right\rangle _{\overline{\phi }}
\end{eqnarray*}
where
\( V'=r_{0}\phi +\lambda _{0}\phi ^{3} \),
\( V''=r_{0}+3\lambda _{0}\phi ^{2} \),
\( \left\langle \bullet \right\rangle _{\overline{\phi }} \) means averaging
on the ensemble with \( \overline{\phi }=\left\langle \phi \right\rangle  \)
fixed, \( r_{0} \) and \( \lambda _{0} \) are, respectively, the quadratic
and quartic coupling.

\section{ICEP}

From eq. (\ref{eq:G_ICEP}) it follows
\begin{eqnarray*}
\Gamma''&=& U''\left( \Omega ,\varphi \right) \\
&& +\frac{1}{2\Omega
}\left[ \frac{U^{iv}\left( \Omega ,\varphi \right) }{U''\left(
\Omega ,\varphi \right) }-\left( \frac{U'''\left( \Omega ,\varphi
\right) }{U''\left( \Omega ,\varphi \right) }\right) ^{2}\right]
\end{eqnarray*}.

The \( U\left( \Omega ,\varphi \right)  \) derivatives involved
above are obtained in a simpler way by suitably exploiting
\cite{myself} eq (\ref{eq:Udef}).

\begin{figure}
\includegraphics[angle=-90,scale=0.38]{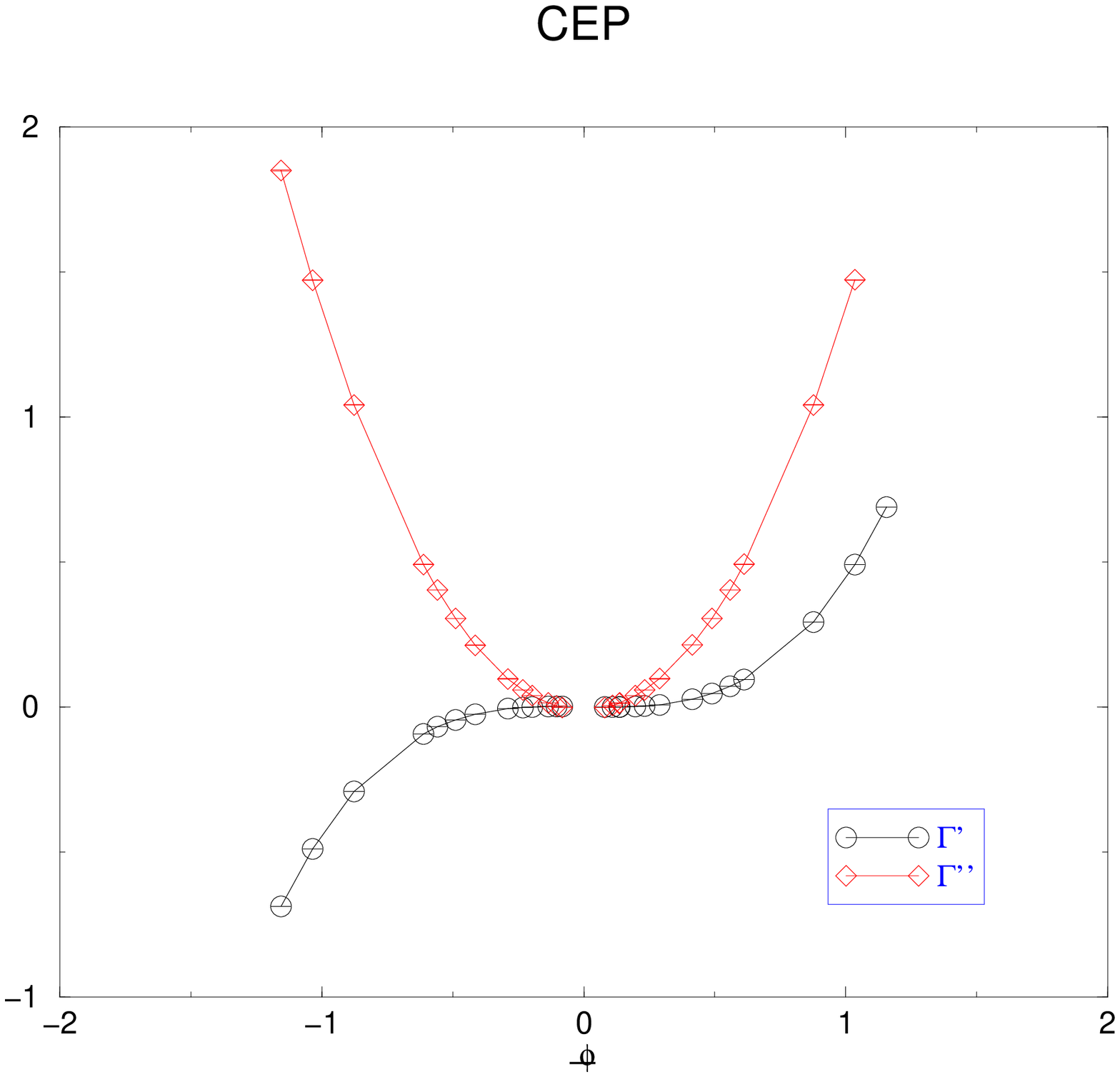}
\caption{Results for $\Gamma'$ and $\Gamma''$ as obtained from CEP}
\end{figure}
\begin{figure}
\includegraphics[angle=-90,scale=0.38]{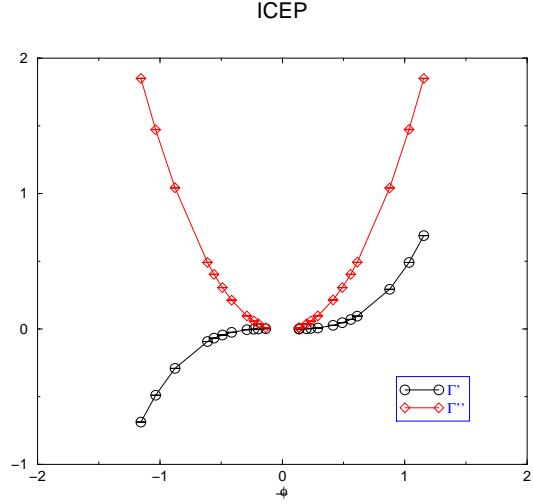}
\caption{Results for $\Gamma'$ and $\Gamma''$ as obtained from ICEP}
\end{figure}

\section{Results}

We have determined $\Gamma'$ and $\Gamma''$ as functions of
$\overline{\phi}$ for $\lambda_0=0.5$, r$_0=-0.2279$ ( near the
critical value),r$_0=-0.2179$ ( symmetric domain ) and
r$_0=-0.2379$ ( broken symmetry domain ).\par With $\varphi$
satisfying $\Gamma'\left(\varphi\right)=0$ one has, by definition, 
$\Gamma''\left(\varphi\right)=m^2$. From our data it turns out
that, for r$_0$ in the symmetric domain  $m^2=0$. Near the
critical value $m^2$ is consistent with a vanishing value. For
r$_0$ in the broken symmetry domain, Fig. 1 shows the CEP results
and Fig. 2 those from ICEP. From these data we get
\vspace{1cm}
\begin{center}
{\centering \begin{tabular}{|c|c|c|}
\hline 
&
\( \varphi  \)&
\( m^{2} \)\\
\hline 
\hline 
CEP&
\( -0.1485\pm 0.0005 \)&
\( 0.0177\pm 0.0002 \)\\
\hline 
&
\( 0.1542\pm 0.0015 \)&
\( 0.0200\pm 0.0006 \)\\
\hline 
ICEP&
\( -0.154\pm 0.001 \)&
\( 0.0163\pm 0.0004 \)\\
\hline 
&
\( 0.155\pm 0.002 \)&
\( 0.0173\pm 0.0008 \)\\
\hline 
\end{tabular}\par}
\end{center}
\vspace{1cm}
\par
The ICEP results are symmetric while the CEP are not. This might be due to ICEP reducing finite size effects.



\begin{thebibliography}{99}
\bibitem[1]{fund2} R. Fukuda and E. Kyryakopoulos, \textit{Nucl. Phys.} \textbf{B85} 354 (1975);
L. O'Raifeartaigh, A. Wipf and H. Yoneyama,  \textit{Nucl. Phys.} \textbf{B271} 653 (1986)
\bibitem[2]{myself} A. Agodi, G. Andronico, P. Cea, M. Consoli and L. Cosmai, \textit{Mod. Phys. Lett.} \textbf{A12}, 1011 (1997);
A. Agodi and G. Andronico, \textit{Nucl. Phys.} \textbf{B(Proc. Suppl.) 73} 730 (1999)
\end{thebibliography}
\end{document}